\documentclass[prb,preprintnumbers,10pt,amsmath,amssymb,a4paper,showpacs,aps,twocolumn]{revtex4-1}
\DeclareMathOperator{\tr}{tr}

\DeclareMathOperator{\spec}{spec}

\providecommand{\eff}[0]{\text{e}}
\providecommand{\mean}[1]{\langle #1 \rangle}
\providecommand{\bra}[1]{\langle #1 |}

\providecommand{\ket}[1]{| #1 \rangle}
\providecommand{\BCS}[0]{| \text{BCS} \rangle}

\providecommand{\fabs}[1]{\left| #1 \right|}

\def\vec#1{\text{\bfseries#1}}

\usepackage{graphicx,color,bm}
\usepackage{subfigure}

\begin{document}

\title{Entanglement spectrum and number fluctuations in the spin-partitioned BCS ground state}
\author{Xavier M. Puspus}
\email{xmpuspus@nip.upd.edu.ph}
\affiliation{National Institute of Physics, University of the Philippines, Diliman, Quezon City 1101, Philippines}
\author{Kristian Hauser Villegas}
\email{kvillegas@nip.upd.edu.ph}
\affiliation{National Institute of Physics, University of the Philippines, Diliman, Quezon City 1101, Philippines}
\author{Francis N. C. Paraan}
\email{fparaan@nip.upd.edu.ph}
\affiliation{National Institute of Physics, University of the Philippines, Diliman, Quezon City 1101, Philippines}

\date{\today}
%\preprint{SanD--0113}

\begin{abstract}
We study entanglement between the spin components of the Bardeen-Cooper-Schrieffer (BCS) ground state by calculating the full entanglement spectrum and the corresponding von Neumann entanglement entropy. The entanglement spectrum is effectively modeled by a generalized Gibbs ensemble (GGE) of non-interacting electrons, which may be approximated by a canonical ensemble at the BCS critical temperature. We further demonstrate that the entanglement entropy is jointly proportional to the pairing energy and to the number of electrons about the Fermi surface (an area law). Furthermore, the entanglement entropy is also proportional to the number fluctuations of either spin component in the BCS state. 
%Furthermore, we demonstrate that the entanglement spectrum approximately corresponds to canonical Gibbs distribution of non-interacting fermions at the BCS critical temperature.
\end{abstract}

%03.67.-a Quantum information
%03.67.Mn QI Entanglement measures, witnesses, and other characterizations 
%03.65.Ud	QM Entanglement and quantum nonlocality 
%75.10.Jm Quantized spin models, including quantum spin frustration
%75.10.Pq	Spin chain models
%74.20.Fg BCS theory

\maketitle
\hbadness=10000

%\section{Introduction}

Bipartite entanglement in a pure state, say $\rho^{AB} = \ket{\psi}\bra{\psi}$, arises from quantum correlations between subsystem partitions $A$ and $B$. Due to these correlations measurements performed on one partition, say $A$, exhibits fluctuations of purely quantum character. Complete information on these subsystem fluctuations is contained in the reduced density operator $\rho^A = \tr_B \rho^{AB}$, which is obtained by averaging over a complete set of states belonging to $B$. Quantifying entanglement in $\rho^{AB}$ involves measuring the degree of uncertainty of the underlying probability distribution over projections onto the Schmidt states of $\rho^A$ (that is, the eigenvalues and eigenvectors of the reduced density operator). A popular scalar measure used for this purpose is the von Neumann entanglement entropy
\begin{equation}
S(\rho^A) = -\tr \rho^A \ln \rho^A,
\end{equation}
which is identical to the Gibbs entropy associated with the probability distribution $\{p_i\} = \spec \rho^A$. Alternatively, the full eigenvalue spectrum of $\rho^A$ may be used as a measure of entanglement in pure states, because comparisons with effective thermal distributions can sometimes provide additional physical insight.\cite{li2008a,schliemann2011a,schliemann2014a}

Many recent studies of entanglement entropy in many-particle systems focus on correlations between spatial partitions.\cite{[{There are several extensive reviews and entire issues on entanglement in many-body systems, for instance, }] amico2008a,*calabrese2009a} {{This emphasis may be based on some current designs of quantum computers that manipulate entangled qubits that are separated in space.\cite{briegel2009a,buluta2011a}}} However, the more general idea of entanglement as a manifestation of quantum correlations makes studies of entanglement under other partitioning schemes valuable in the understanding of interacting systems. For instance, a general scheme for the computation of modewise entanglement entropy that is relevant to the system discussed here has been derived for bosonic\cite{botero2003a,rendell2005a,cramer2006a,ciancio2006a} and fermionic\cite{botero2004a,kraus2009a} Gaussian states. One of the main conclusions in these papers is that the analysis of mode entanglement in such Gaussian states can be reduced to an analysis of two-mode (pair-wise) entanglement, which greatly simplifies the theoretical study of entanglement in these many-body systems. {Also, mode entanglement has been studied previously in the context of examining single-particle nonlocal quantum effects (Bell inequalities)\cite{tan1991a, hardy1994a, hessmo2004a,ashhab2009a,vanenk2005a} and extractable entanglement from assemblies of identical particles for quantum information processing tasks (entanglement of particles).\cite{wiseman2003a,dowling2006a, bartlett2006a,ashhab2007a,ashhab2007b,friis2013a} }

In this paper, we calculate the entanglement entropy present between the spin components of an electron system with pair interactions. That is, we partition the ground state of a mean-field Bardeen-Cooper-Schrieffer (BCS) model\cite{bcs1957} into spin-up and spin-down subsystems and compute the von Neumann entropy in the resulting reduced state. We refer to this entropy as the spin entanglement entropy or spin-EE to emphasize the  chosen partitioning scheme. Previous studies of quantum correlations and entanglement in the BCS state under similar mode partitioning have used different measures such as concurrence,\cite{martindelgado2002,dunning2005a,oh2005a,gao2007a} negativity,\cite{chung2008a} and pairing.\cite{kraus2009a} Close to our work is a general calculation for the modewise entanglement in a pure Gaussian state,\cite{botero2004a} of which the BCS ground state is an example. The work presented here is different from these in two important respects. First, as opposed to the local entanglement measures (in momentum space) reported,\cite{oh2005a,gao2007a} we investigate entanglement in the full many-particle ground state to establish an area law.\cite{eisert2010a} Second, our use of the entanglement spectrum and von Neumann entropy as measures allows us to obtain simple results that clarify the physical relationships between component interactions, entanglement entropy, and number fluctuations. In particular, we establish the analytical dependence of the entanglement spectrum and spin-EE on the physical parameters of the BCS model: the pairing energy $\Delta$ (which depends on the electron-phonon coupling strength) and the density of single particle orbitals $g(0)$ at the Fermi energy $\mu$ (which depends on the mean number of {electrons in Cooper pairs}). {{These contributions may prove useful in the study of entangled Cooper pairs in the wake of recent proposals to extract them from superconducting tips via field emission into vacuum.\cite{yuasa2009a,giovannetti2012a} Additionally, the results presented here allow one to make general statements about the scaling laws obeyed by these quantities in BCS and BCS-like states and provide a practical means of quantifying entanglement by measuring ground state fluctuations.\cite{[{The relationship between entanglement measures and quantum metrology techniques are discussed, for instance, in }] benatti2010a,*benatti2014a} Furthermore, we demonstrate that an analysis of the entanglement spectrum of these ground states yields information on the critical properties of the model at finite temperature.}}

Our study begins with a brief review of the BCS model, with an emphasis on essential features that are relevant to the generation of ground state spin entanglement (Sec.~\ref{sect:model}). We then construct a thermal model of non-interacting fermions that allow us to treat the statistical effects of pair formation and annihilation in the reduced single component state as resulting from effective thermal excitations (Sec.~\ref{sect:spec}). Next, we calculate the total spin entanglement entropy in the BCS state and discuss its simple relationship with the pairing energy, the number of electrons forming Cooper pairs, and the number fluctuations in the ground state (Sec.~\ref{sect:ee}). Finally, we point out that these results also apply to mean-field models with bilinear intercomponent coupling (Sec.~\ref{sect:gen}).

\section{Model}\label{sect:model}

%\textit{Model}---
We consider here the model BCS hamiltonian
\begin{equation}
%H_\text{mf} = \sum_{\vec{k}\sigma} \xi_\vec{k} c_{\vec{k}\sigma}^\dagger c_{\vec{k}\sigma}
%-\sum_{\vec{k}}
%\bigl(\Delta_\vec{k}c_{\vec{k}\uparrow}^\dagger c_{-\vec{k}\downarrow}^\dagger +   
%\Delta_\vec{k}^* c_{-\vec{k}\downarrow}c_{\vec{k}\uparrow}\bigr)
H = \sum_{\vec{k}\sigma} \xi_\vec{k} c_{\vec{k}\sigma}^\dagger c_{\vec{k}\sigma}
-\Delta\sum_{\vec{k}}'
\bigl(c_{\vec{k}\uparrow}^\dagger c_{-\vec{k}\downarrow}^\dagger +   
c_{-\vec{k}\downarrow}c_{\vec{k}\uparrow}\bigr)
\end{equation}
The electron orbital energy $\xi_\vec{k} = \epsilon_\vec{k} - \mu$ is measured with respect to the Fermi energy $\mu$ and the pairing energy $\Delta$ is approximated to be independent of electron wavevector. The prime in the second sum means that only electrons with energy within the Debye shell $\xi_\vec{k} \in [- \epsilon_\text{D},\epsilon_\text{D}]$ interact attractively to form Cooper pairs (the Debye energy $\epsilon_\text{D}$ is the phonon energy scale). The mean-field hamiltonian $H$ is bilinear in fermion operators and therefore its eigenstates are fermionic Gaussian states.\cite{botero2004a,kraus2009a} Entanglement measures in these states can therefore be calculated from the reduced correlation functions of the model\cite{botero2003a,peschel2003a} or from the exact diagonalization of the reduced density operator, as we do below.

The ground state of the hamiltonian $H$ is the BCS wavefunction
\begin{equation}\label{eq:bcsstate}
\BCS = \bigotimes_\vec{k}\, \bigl(u_{\vec{k}} + v_{\vec{k}} c^\dagger_{\vec{k}\uparrow} c^\dagger_{-\vec{k} \downarrow}\bigr) | 0 0 \rangle_\vec{k}.
\end{equation}
This is a linear superposition of all possible occupancies of Cooper $\vec{k}$-pairs $\ket{n_{\vec{k}\uparrow}n_{-\vec{k}\downarrow} }_\vec{k}$, where $u_\vec{k}$ ($v_\vec{k}$) is the probability amplitude for the $\vec{k}$-pair orbital being unoccupied (occupied). Inside the Debye shell
\begin{align}
 \fabs{u_\vec{k}}^2&\equiv \fabs{u(\xi_\vec{k})}^2 = \frac{1}{2}\biggl[1 + \frac{\xi_\vec{k}}{\sqrt{\xi_\vec{k}^2 + \Delta^2}} \biggr], \\
 \fabs{v_\vec{k}}^2&\equiv \fabs{v(\xi_\vec{k})}^2 = 1 - \fabs{u_\vec{k}}^2,
\end{align}
while outside this shell $\fabs{v_\vec{k}}^2 = 1$ ($\fabs{v_\vec{k}}^2 = 0$) for $\xi_\vec{k}<-\epsilon_\text{D}$ ($\xi_\vec{k}>+\epsilon_\text{D}$). The quantity $\fabs{v_\vec{k}}^2$ is therefore the probability that the pair orbital $\ket{n_{\vec{k}\uparrow}n_{-\vec{k}\downarrow} }_\vec{k}$ is occupied.  

The orbital $\ket{n_{\vec{k}\uparrow}n_{-\vec{k}\downarrow} }_\vec{k}$ in eq.~\eqref{eq:bcsstate} is labeled by the wavevector of the spin-up electron of the Cooper pair. In this form, the BCS wavefunction is manifestly Schmidt decomposed with respect to the different spin components. This observation is important because it implies that the reduced density operator $\rho^\uparrow \equiv \tr_\downarrow \rho^{\uparrow\downarrow}$ is diagonal in the Fock basis of spin-up electron orbital occupancies. This fact greatly simplifies the analysis of spin entanglement in the $\BCS$ state.

The full density operator $\rho^{\uparrow\downarrow} = \BCS \bra{\text{BCS}}$ for the BCS ground state is
\begin{align}
\rho^{\uparrow\downarrow} = \bigotimes_{\vec{k}\vec{k}'}&\bigl( u_\vec{k}u_{\vec{k}'}^* |00\rangle_\vec{k} \langle 00|_{\vec{k}'} 
+ v_\vec{k}v_{\vec{k}'}^* |11\rangle_\vec{k} \langle 11|_{\vec{k}'} \nonumber \\ 
&\ + u_{\vec{k}} v_{\vec{k}'}^* |00\rangle_\vec{k}\langle 11|_{\vec{k}'} 
+ v_{\vec{k}} u_{\vec{k}'}^* |11\rangle_\vec{k}\langle 00|_{\vec{k}'}\bigr).
\end{align}
Averaging over all possible occupancies of spin-down electrons gives the reduced density operator for the spin-up electrons
\begin{equation}
\rho^\uparrow = \tr_{\downarrow} \rho^{\uparrow\downarrow} = \bigotimes_\vec{k}\bigl(|u_\vec{k}|^2 |0\rangle_\vec{k} \langle 0|_\vec{k} + |v_\vec{k}|^2 |1\rangle_\vec{k} \langle 1|_{\vec{k}}\bigr).
\end{equation}
This reduced density operator $\rho^\uparrow =\bigotimes_\vec{k} \rho^\uparrow_\vec{k}$ acts on a tensor product space that consists of the independent state spaces of spin-up electrons, each labeled by the wavevector $\vec{k}$.

\begin{figure}[tb]
	\centering
		\includegraphics[width=0.85\linewidth]{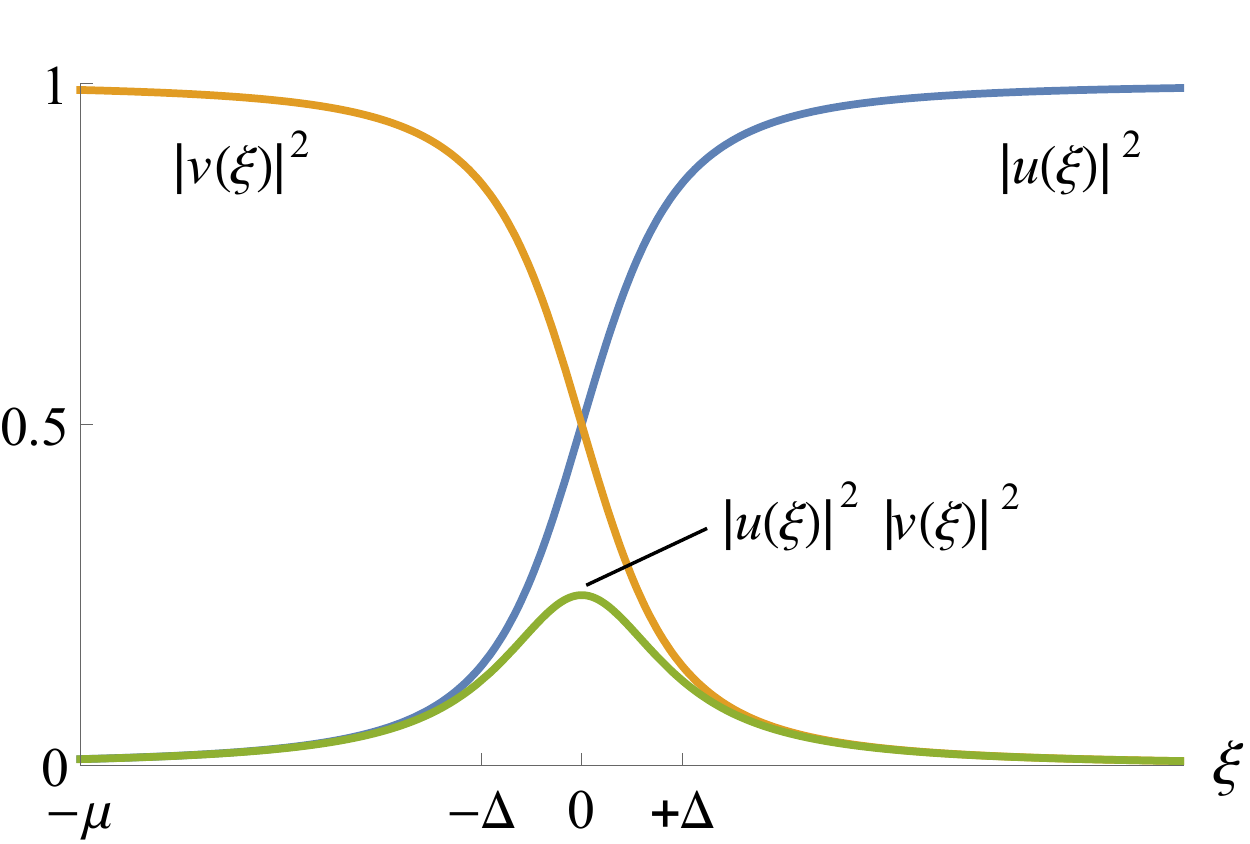}
	\caption{(Color online) Entanglement spectrum of the spin-partitioned BCS ground state. Discontinuities at $\pm \epsilon_\text{D}$ have been smoothed out.}
	\label{fig:spectrum}
\end{figure}

\section{Entanglement spectrum and effective thermal model}\label{sect:spec}

%\textit{Entanglement spectrum and effective thermal model}.---
The entanglement spectrum $\spec \rho^\uparrow$ consists of the set of all probabilities $\{\fabs{u_\vec{k}}^2,\left|v_\vec{k}\right|^2\}$ (Fig.~\ref{fig:spectrum}). This spectrum is qualitatively similar to the probability distribution of orbital occupancies of non-interacting fermions at thermal equilibrium (unit occupancy deep within the Fermi surface, zero occupancy far outside it, and a smooth transition of width $\Delta\sim \beta^{-1}$).\cite{poole1995,*tinkham1996} The mapping between the entanglement spectrum and an effective thermal distribution of orbital occupancies may be accomplished by the method of correlation functions.\cite{peschel2003a} To do so, we require 
\begin{equation}\label{fermidist}
 \tr \rho^\uparrow c_{\vec{k}\uparrow}^\dag c_{\vec{k}\uparrow} = \fabs{v(\xi_\vec{k})}^2 = \frac{1}{1+e^{\beta_\eff \xi_\vec{k}}},
%e^{-\beta_\eff\xi_\vec{k}} \longleftrightarrow \frac{\abs{v_\vec{k}}^2}{\abs{u_\vec{k}}^2}= \frac{\sqrt{\xi_\vec{k}^2 +\Delta^2} - \xi_\vec{k}}{\sqrt{\xi_\vec{k}^2+\Delta^2} + \xi_\vec{k}}.
\end{equation}
which gives the effective reciprocal temperature
\begin{equation}\label{beffexact}
	\beta_\eff(\xi_\vec{k}) = \frac{2}{\xi_\vec{k}} \coth^{-1}\frac{\sqrt{\xi_\vec{k}^2+\Delta^2}}{\xi_\vec{k}}.
	%e^{-\beta_\eff\xi_\vec{k}} \longleftrightarrow \frac{\abs{v_\vec{k}}^2}{\abs{u_\vec{k}}^2}= \frac{\sqrt{\xi_\vec{k}^2 +\Delta^2} - \xi_\vec{k}}{\sqrt{\xi_\vec{k}^2+\Delta^2} + \xi_\vec{k}}.
\end{equation}
It turns out that the exact reciprocal temperature $\beta_\eff$ is a function of orbital momentum and thus the effective thermal analog of the reduced state $\rho^\uparrow$ is a generalized Gibbs ensemble (GGE)\cite{rigol2007b} of non-interacting spin-polarized fermions:
\begin{equation}
	\rho_\text{e}^\uparrow = \frac{e^{-\sum \beta_\text{e}(\xi_\vec{k}) \xi_\vec{k} c_{\vec{k}\uparrow}^\dag c_{\vec{k}\uparrow}}}{\tr e^{-\sum \beta_\text{e}(\xi_\vec{k}) \xi_\vec{k} c_{\vec{k}\uparrow}^\dag c_{\vec{k}\uparrow}}}.
\end{equation}
As the pairing energy $\Delta$ goes to zero, the effective temperature also goes to zero for all $\vec{k}$, which yields a reduced state that is consistent with the ground state of an electron gas.

\begin{figure}[tb]
	\centering
		\includegraphics[width=0.8\linewidth]{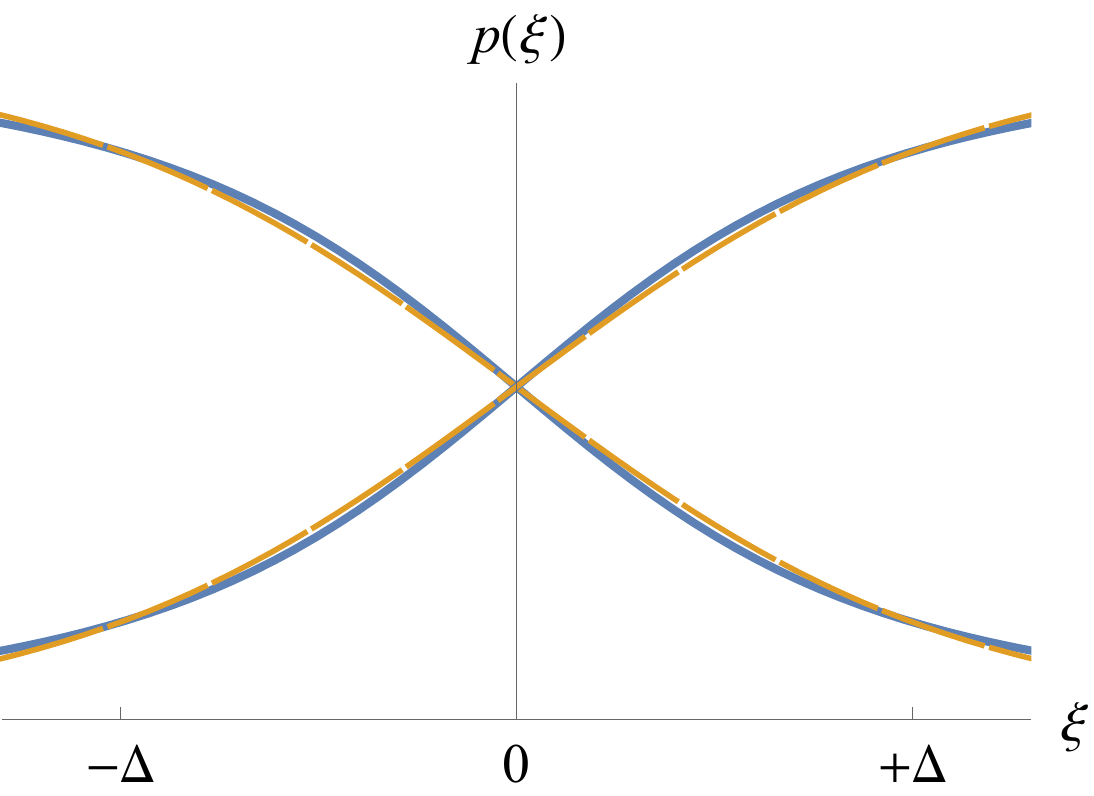}
	\caption{(Color online) The entanglement spectrum $\spec \rho^\uparrow$ (solid line) is comparable to the orbital occupancies of an effective thermal theory of non-interacting fermions at reciprocal temperature $\beta_\eff^0 \approx 1.76/\Delta$ (dashed line). For the latter $p(\xi) = 1- (1+e^{\beta_\eff^0 \xi})^{-1}$ and $(1+e^{\beta_\eff^0 \xi})^{-1}$.}
	\label{fig:weights}
\end{figure}

To obtain an approximate effective canonical Gibbs ensemble (constant temperature), we can define a constant $\beta_\eff^0$ for which the mapping \eqref{fermidist} holds identically at $\xi_\vec{k} = \pm\Delta$. Doing so leads to the constant reciprocal temperature
\begin{equation}\label{beff}
\beta_\eff^0 \equiv \frac{2}{\Delta} \coth^{-1}\sqrt{2} = \frac{1}{\Delta} \ln \biggl(\frac{\sqrt{2} + 1}{\sqrt{2} -1} \biggr) \approx \frac{1.7627}{\Delta}.
\end{equation}
The approximation $\beta_\eff(\xi) \to \beta_\eff^0$ is good in the vicinity of the Fermi surface $\xi_\vec{k}\in [-\Delta,\Delta]$ where entanglement is greatest $\left|{u_\vec{k}}\right|^2 \approx \left|{v_\vec{k}}\right|^2$ (Fig.~\ref{fig:weights}).

%Unfortunately, this mapping does not have a solution for a reciprocal temperature $\beta_\eff$ independent of $\xi$. 
We remark that the effective temperature $\beta_\eff^0$ describing the entanglement spectrum of either spin component is approximately equal to the {BCS critical temperature\cite{poole1995,*tinkham1996} 
\begin{equation}\label{bcrit}
\beta_c = \frac{\pi e^{-\gamma}}{\Delta} \approx \frac{1.7639}{\Delta} \approx \beta_\eff^0,	
\end{equation}
$\gamma$ being the Euler-Mascheroni constant. This correspondence suggests that the difference between the entanglement entropy of the BCS ground state and the unentangled normal metal state may be physically interpreted as a measure analogous to the difference in free energy between the superconducting and normal phases. As an aside, since the integrals used to evaluate the effective \eqref{beff} and critical \eqref{bcrit} temperatures are distinct, their approximate equality may be of interest to number theorists.}

%Finally, we notice the approximate relation 
%\begin{equation}
%\frac{\pi}{ e^{\gamma} }\approx \ln \biggl(\frac{\sqrt{2} + 1}{\sqrt{2} -1} \biggr) \approx 1.76,
%\end{equation}
%which may be of interest to number theorists.

\section{Entanglement entropy}\label{sect:ee}
\begin{figure}[tb]
	\centering
		\includegraphics[width=0.9\linewidth]{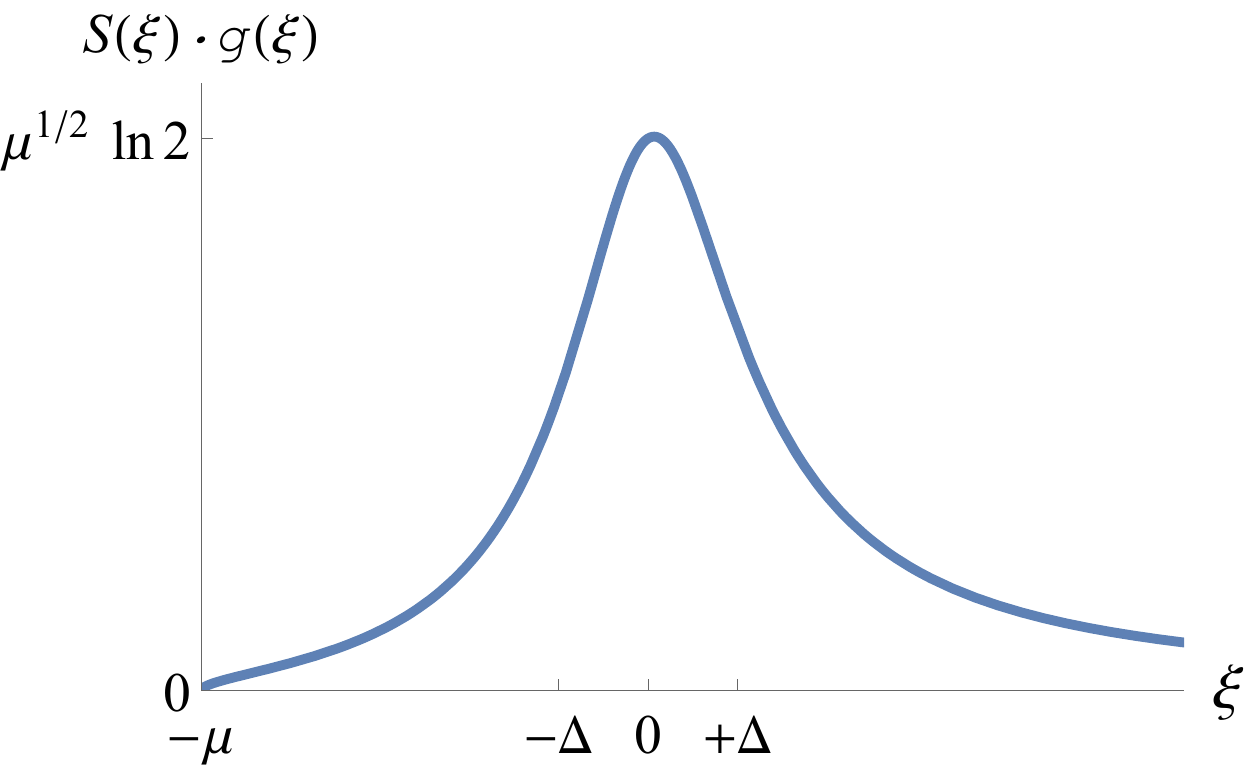}
	\caption{(Color online) Density of states weighted entanglement entropy of energy orbitals. The density of states used here $g(\xi) = (\xi+\mu)^{1/2}$ corresponds to a 3D dispersion relation quadratic in momentum. The total spin-EE is dominated by the contributions of orbitals in the vicinity of the Fermi energy.}
	\label{fig:weighted}
\end{figure}

%\textit{Entanglement entropy}.---
The total spin entanglement entropy in the BCS state $S^\uparrow = -\tr \rho^\uparrow \ln \rho^\uparrow$ is a sum of partial contributions $0\le S^\uparrow_\vec{k} \le \ln 2$ from each $\vec{k}$-pair orbital in the Debye shell:
\begin{equation}
S^\uparrow = \sum_{\vec{k}}' S^\uparrow_\vec{k} = -\sum_{\vec{k}}' \tr \rho_\vec{k}^\uparrow \ln \rho_\vec{k}^\uparrow.
\end{equation}
In the thermodynamic limit, the spin-EE is given by the integral
\begin{equation}
S^\uparrow = \int_{-\epsilon_\text{D}}^{\epsilon_\text{D}} S(\xi) g(\xi) \,d\xi,
\end{equation}
where $S(\xi)=-\bigl[\fabs{u(\xi)}^2\ln \fabs{u(\xi)}^2 + \fabs{v(\xi)}^2\ln \fabs{v(\xi)}^2\bigr]$ and the density of states $g(\xi)$ can be calculated from the bare dispersion relation $\xi_\vec{k}$.\cite{braunstein2005a}  The total spin-EE can be calculated exactly when the pairing energy is much smaller than the Debye and Fermi energies, $\Delta \ll \epsilon_\text{D} \ll \mu$, so that $g(\xi)\approx g(0)$ within the Debye shell and the limits of the integral can be extended to $\pm\infty$. This approximation is justified by the fact that the partial entanglement entropy $S(\xi)$ is peaked about the Fermi surface with a width of the order of $\Delta$. Evaluating the integral gives
\begin{equation} \label{spin-ee}
S^\uparrow = \pi g(0) \Delta.
\end{equation}
The quantity $g(0)\Delta $ is the approximate number of electron orbitals in a shell of width $\sim 2\Delta$ about the Fermi energy $\mu$, which is precisely the interaction region where partial contributions to the total spin-EE are largest (Fig.~\ref{fig:weighted}). Additionally, for $\Delta\ll \mu$ the entanglement entropy $S^\uparrow$ is proportional to the number of orbitals on the Fermi surface and we have an ``area'' law\cite{eisert2010a} for entanglement entropy between the two spin sectors. The area law \eqref{spin-ee} holds regardless of the dimensionality of the model; The interactions between spin components are short-ranged or ``local'' in the sense that an electron with momentum and spin $\vec{k}\negmedspace\uparrow$ interacts with only one other electron $-\vec{k}\negmedspace\downarrow$. We remark that this example of component entanglement due to Cooper pairing within an energy shell has similarities to the valence bond entanglement entropy, which is a measure of the number of spin-singlets shared between spatial partitions in a valence bond state.\cite{alet2007a} This analogy clarifies the picture of entanglement as a measure of correlations across a partition and the general expectation of area laws when correlations are short-ranged.

The simple result \eqref{spin-ee} shows that the spin-EE is proportional to the pairing energy $\Delta$. This is a clear demonstration of how interactions between spin components lead to entanglement in a many-body system. When the coupling between spin-up and spin-down electrons vanishes at the superconductor-normal metal transition ($\Delta\to 0$) the ground state mode entanglement vanishes also. {Indeed, a similar measure of total mode entanglement in the BCS ground state called the macrocanonical entanglement of pairing (MEP)\cite{martindelgado2002} has been shown to vanish as $\Delta \to 0$. The MEP is a measure based on the product of partial concurrences and has similar properties to the total spin-EE considered here.}\footnote{In fact, with the approximation $\Delta\ll \epsilon_\text{D}$ it turns out that $S^{\uparrow} = -2\ln (1- \text{MEP})$. The value of using the von Neumann entropy here is that the area law eq.~\eqref{spin-ee} is manifest.} The fact that these entanglement measures vanish when the pairing energy vanishes agrees completely with our previous result \eqref{beff} that the temperature of the effective thermal model matches the critical BCS temperature.

A seemingly contrasting result\cite{subrahmanyam2010a} emphasizes the importance of properly defining the partitioning scheme when discussing entanglement entropy. In a previous study of a spatially partitioned BCS ground state on a lattice, the single-site entanglement entropy shows a completely opposite dependence on the pairing energy versus the spin-EE.\cite{subrahmanyam2010a} It turns out that the single-site entropy is maximum in the normal metal $\Delta=0$ and diminishes as the gap opens in the superconducting phase. 

\subsection*{Number fluctuations}
%{\it Number fluctuations.}---

The reduced spin-up electron state $\rho_\uparrow$ is a statistical operator over Fock states and the {{number of paired spin-up electrons $N_\uparrow$}} therefore fluctuates in the ground state. {{The associated variance in spin-up electron number is equal to
\begin{equation}
	\sigma_\uparrow^2 = \tr \bigl[({N}_\uparrow - \mean{{N}_\uparrow})^2 \rho_\uparrow\bigr] = \sum_\vec{k} \fabs{u_\vec{k}}^2 \fabs{v_\vec{k}}^2.
\end{equation}
In the same approximation $\Delta \ll \epsilon_\text{D} \ll \mu$ used to calculate $S_\uparrow$, we find
\begin{equation}
	\sigma_\uparrow^2 \approx \tfrac{1}{4}\pi g(0)\Delta,
\end{equation}
which is one-fourth of the variance in total electron number in $\BCS$, $\sigma_{\uparrow\downarrow}^2 = \pi g(0)\Delta$. This relationship can be interpreted physically as follows: Pair interactions controlled by $\Delta$ lead to pair number fluctuations in the BCS ground state. Thus, electrons in the BCS state are gained and lost in pairs of opposite spin. This correlated fluctuation in the number of electrons of opposite spin gives rise to uncertainty in the determination of the reduced states $\rho_\uparrow$ and $\rho_\downarrow$ and, hence, to non-zero entanglement entropy. {In fact, it turns out that the spin-EE and the number fluctuations in the BCS state are equal: $S_\uparrow = \sigma_{\uparrow\downarrow}^2 = 4 \sigma_\uparrow^2$. A similar relationship has been reported between local orbital concurrence and occupation number fluctuations.\cite{dunning2005a} These numerical equivalences can be expected from a mean-field theory with BCS-like interactions and has been anticipated by the exact calculation of the full counting statistics (FCS) function for the BCS state.\cite{belzig2007}} 

The proportionality between entanglement entropy and subsystem number fluctuations has been explored previously in spatially-partitioned fermion gases and other conformally-invariant theories.\cite{song2010a,calabrese2012a,song2012a} However, the mechanism behind the number fluctuations in these latter examples is the gain/loss of particles from one spatial partition to another, and not the population/depopulation of electron orbitals as in the spin-partitioned BCS state discussed here. }}

\section{Generalizations}\label{sect:gen}
%{\it Generalizations}---
Entanglement between the spin components is unaffected by unitary transformations within the subspace of either component. Our results therefore apply to ground states of other models with bilinear interactions between components. For example, a particle-hole transformation in one component gives the interaction term $\Delta\bigl(c_{\vec{k}\uparrow}^\dagger c_{-\vec{k}\downarrow} +   
c_{-\vec{k}\downarrow}^\dagger c_{\vec{k}\uparrow}\bigr)$, which can describe elastic scattering of spin-flipped electrons by fixed magnetic impurities in a mean-field approach.\cite{ma2004a} Furthermore, the spin labels may be replaced by other component or isospin labels and our results may be used to describe mode entanglement in other two component systems such as electron system bilayers.\cite{schliemann2011a,schliemann2013a,lee2014a} The main requirement for the validity of these generalizations is that the component interactions are quasi-local in momentum space, as they are in the mean-field theory of Cooper pairing.

Additionally, since the total spin-EE is a sum over independent partial terms, one can expect a scaling relationship similar to eq.~\eqref{spin-ee} to hold even when the pairing energy $\Delta$ depends on the wavevector $\vec{k}$, as long as $\Delta_\vec{k}$ slowly varies within the Debye shell.

\section*{Concluding remarks}
%{\it Concluding remarks.}---

We have completely characterized the entanglement between spin components of the BCS ground state. The reduced states are mixed and described by a probability distribution of occupancies that is identical to a generalized Gibbs distribution with momentum dependent temperature, and similar to that of an effective non-interacting fermion gas at thermal equilibrium. We demonstrated that the temperature of the effective canonical ensemble is nearly equal to the critical temperature $\beta_\eff^0 = \Delta^{-1}\ln[(2^{1/2}+1)/(2^{1/2}-1)] \approx 1.76/\Delta$. {We emphasize that this result, which uses an effective thermal description of the partitioned ground state ($T = 0$), is distinct from (but related to) the well-established critical value $\beta_c = \pi e^{-\gamma}/\Delta$. The latter result is derived from the thermodynamics of quasiparticle excitations at non-zero temperature ($T>0$).} Also, we have calculated the spin entanglement entropy in $\BCS$ and showed that it obeys an area law that is independent of system dimensionality \eqref{spin-ee}. That is, the spin-EE is proportional to the number of electron orbitals about the Fermi surface. Finally, we provided exact quantitative arguments that the spin-EE in BCS and BCS-like ground states arise from ground state fluctuations in the occupancy of Cooper pair orbitals.

In general, the entanglement entropy is a measure of the number of correlated degrees of freedom across a partition weighted by the strength of correlations. The simple example provided by the BCS ground state illustrates this in a very simple manner. The results presented here are not sensitive to the particular model under consideration in the following sense. If the effective distribution of electron (or fermionic quasiparticle) occupancies obtained from the entanglement spectrum possesses the following qualitative features: (i) unit occupation deep within the Fermi sphere, (ii) zero occupation far above the Fermi sphere, and (iii) a smooth transition of width $\sim\negthinspace\Delta$ about the Fermi energy, then the component entanglement entropy will scale according to the area law $\sim  \pi g(0) \Delta$.

\begin{acknowledgments}
This work is supported by the University of the Philippines OVPAA through Grant No.~OVPAA-BPhD-2012-05. The authors acknowledge helpful conversations with Shikano Y., Katsura H., W.~Belzig, M.~A.~Mart{\'{i}}n-Delgado, and U.~Marzolino.
\end{acknowledgments}

%\bibliography{seebcsbib}

\end{document}